\documentclass[prd,nofootinbib,superscriptaddress,preprint]{revtex4}
\usepackage[T1]{fontenc}
\usepackage{amsmath,amssymb}
\usepackage{epsfig}
\usepackage{graphicx,array}
\usepackage[usenames,dvipsnames]{color}
\usepackage{slashed}
\usepackage{comment}
\usepackage[colorlinks,citecolor=blue]{hyperref}
\usepackage{pdfpages,orcidlink}
\usepackage{color}
\usepackage{subcaption}

\usepackage{multirow}

\begin{document}

	\title{ From Dirac Neutrino Mass to Dark Radiation: An $A_4$-Symmetric FIMP Dark Matter Framework} 
	\author{Pritam Das$^{\orcidlink{0000-0001-6195-6944}}$}
	\email{prtmdas9@gmail.com}
	\affiliation{Department of Physics, Salbari College, Baksa, Assam 781318, India}
	
	\author{Anupam Deka$^{\orcidlink{0009-0001-6548-2659}}$}
	\email{anuamdeka585@gmail.com}
	\affiliation{Department of Physics, Nabajyoti College,Barpeta,Assam 781319, India}
	\affiliation{Department of Physics, Bhattadev University, Bajali, Assam 781325, India}

	\author{Debajyoti Dutta$^{\orcidlink{0000-0002-8348-6098}}$}
	\email{phy.debajyoti@bhattadevuniversity.ac.in}
	\affiliation{Department of Physics, Bhattadev University, Bajali, Assam 781325, India}

	\begin{abstract}
		We propose a unified $A_4\times Z_2\times U(1)_L$ framework for generating small Dirac neutrino masses and address dark matter through feeble interactions in an extended particle physics model. The model extends the Standard Model with  a pair of heavy vector-like neutral $A_4$ fermions $N(N_{L,R})$, two additional flavon fields $\eta$ and $\phi$, among which $\eta$ being a $Z_2$-odd inert scalar and a dark-sector vector-like fermion field $\chi$. 
		We numerically explore the Dirac-sector Yukawa couplings and identify the parameter region that is consistent with the {\tt NuFit 6.1} neutrino oscillation data at the $3\sigma$ level. The dark matter candidate $\chi$, stabilized by the $Z_2$ symmetry, is produced non-thermally through the decay $N_L\rightarrow\chi+\eta$. By solving the Boltzmann equation, we find that the observed relic abundance, $\Omega_{\rm DM}h^2\simeq0.12$, can be reproduced for freeze-in couplings of order $\lambda\sim10^{-5.5}$--$10^{-7.5}$ over the considered dark matter mass range. For the chosen parameter space, we also verify the interaction rate remains below the Hubble expansion rate, confirming  $\chi$ never reaches thermal equilibrium. Thus, our results demonstrate that the model provides a consistent realization of Dirac neutrino masses and FIMP dark matter within a common symmetry-based setup. The presence of light Dirac RH neutrinos can contribute to the dark radiation and hence the effective degrees of freedom via thermal and non-thermal processes. We further estimate the contribution of the light right-handed neutrinos to \(\Delta N_{\rm eff}\), finding a maximum characteristic value of \(\Delta N_{\rm eff}\simeq0.14\) in the benchmark parameter region.
	\end{abstract}

	
	\maketitle 
	
	\section{Introduction}\label{sec:intro} 
	The discovery of neutrino oscillations established the first evidence of physics beyond the Standard Model (SM) \cite{Super-Kamiokande:2001bfk,T2K:2011ypd,KamLAND:2008dgz,DoubleChooz:2011ymz,DayaBay:2012fng,RENO:2012mkc,MINOS:2013xrl}, confirming that neutrinos possess tiny but non-zero masses and flavor mixing. Since neutrinos are massless in the SM, to address their masses theoretically, one requires an extension beyond the SM. In connection to the neutrino mass, several fundamental questions remain open, including whether neutrinos are Dirac or Majorana particles and the nature of the neutrino mass ordering. Although collider searches \cite{CMS:2018jxx,Cai:2017mow} and neutrinoless double-beta decay (0$\nu\beta\beta$) experiments \cite{Dolinski:2019nrj,EXO-200:2019rkq,LEGEND:2021bnm,CUPID:2020aow,KamLAND-Zen:2022tow} have found no evidence for lepton-number violation. While the negative results from these experiments themselves do not establish the Dirac nature of neutrinos, there has been an increasing interest in Dirac neutrino models. Dirac neutrino models have received considerable attention, with various mechanisms generating tiny Dirac masses through tree or loop-level seesaw frameworks and extended scalar sectors \cite{Minkowski:1977sc, GellMann:1980vs,Mohapatra:1979ia,Sawada:1979dis,Yanagida:1980xy,Schechter:1980gr, Lazarides:1980nt, Wetterich:1981bx, Schechter:1981cv, Foot:1988aq,Borah:2024yow,Borah:2024gql,Mohapatra:1980de}. Such constructions often employ additional discrete or continuous symmetries, such as $A_4$, $Z_2$, and $U(1)_{B-L}$, to generate the observed neutrino masses while forbidding unwanted Majorana terms \cite{Chulia:2016ngi,Bonilla:2016zef,Borah:2017dmk,CentellesChulia:2018gwr,CentellesChulia:2018bkz,Han:2018zcn,Borah:2018gjk,Borah:2018nvu,Borah:2019bdi,Jana:2019mgj,Nanda:2019nqy,Guo:2020qin,Bernal:2021ezl,Borah:2022obi,Das:2026xjq}.

	Another puzzling issue in modern particle physics is the existence of dark matter (DM), which constitutes $26.8\%$ of the known universe. The presence of DM is supported by various astrophysics observations along with cosmological experiments like Planck~\cite{Planck:2018vyg} and WMAP~\cite{ParticleDataGroup:2020ssz}. The latest cosmological observation put the relic abundance of DM to be $\Omega_{\mathrm{DM}} h^2 = 0.120 \pm 0.001$~\cite{Planck:2018vyg}. Although none of the SM particles can explain the origin of DM, weakly interacting massive particle (WIMP) has emerged as the leading candidate to explain the DM abundance via thermal freeze-out. In the WIMP scenario, the DM is initially in thermal equilibrium with the primordial plasma and subsequently freezes out when its annihilation rate becomes smaller than the expansion rate of the universe. However, the null results at the direct detection of DM experiments have motivated the exploration of alternative scenarios.
	
	A compelling alternative to the traditional thermal freeze-out scenario is provided by the freeze-in mechanism \cite{Hall:2009bx,Bernal:2017kge}, in which dark matter (DM) belongs to the class of feebly interacting massive particles (FIMP). Due to their extremely suppressed interactions, FIMP candidates never reach the thermal equilibrium with the Standard Model (SM) thermal bath \cite{Hall:2009bx,Blennow:2013jba,Klasen:2013ypa}. Instead, they are gradually produced via the decays and scatterings of heavy in-equilibrium particles in the early Universe \cite{Hall:2009bx,Elahi:2014fsa}. Because the production rate is much slower than the expansion rate of the Universe, the DM yield grows cumulatively until the temperature drops below the mass threshold of the parent particles (or the DM mass itself). Then at that point production effectively ceases and the DM abundance freezes in \cite{Hall:2009bx,Bernal:2017kge}. The FIMP paradigm naturally accounts for the observed DM relic density while simultaneously explaining the non-observation of DM in direct, indirect, and collider searches \cite{Bernal:2017kge}.
	
	Motivated by this, we propose a Dirac-seesaw mechanism for production of the Dirac neutrino mass and simultaneous production of DM in this work. We consider the non-Abelian discrete flavor symmetry group $A_4$ \cite{Ma:2001dn,Babu:2002dz,Altarelli:2005yp,Altarelli:2005yx,Altarelli:2010gt} with an additional discrete symmetry group $Z_2$ and global lepton number symmetry $U(1)_L$. In addition to the particle content of SM, it contains additional light right handed neutrinos ($\nu_R$), a pair of heavy vector-like neutral fermion $N(=N_L,N_R)$, vector-like neutral fermion ($\chi$) and additional flavon field ($\eta$). The $A_4$ symmetry spontaneously gets broken by the flavon fields, which generate neutrino masses and mixing through a seesaw mechanism \cite{Borah:2017dmk}. The dark sector consists of additional fields $\chi$ and $\eta$ both are odd under imposed $Z_2$ symmetry and they are produced from the decay of heavy fields through the decay process $N_L\rightarrow \chi+\eta$. The lightest of $Z_2$ particles will remain stable and act as a DM candidate \cite{Borah:2018gjk,Becker:2018rve,Biswas:2022vkq}.

	A natural consequence of the Dirac nature of light neutrinos is the existence of three light right-handed states, $\nu_R$. The presence of these light states can contribute to the energy density hence $N_{\rm eff} $ buildup, either via thermally or non-thermally, depending on their interaction strength~\cite{Luo:2020sho,Luo:2020fdt, Borah:2022enh, Borah:2025fkd}. Cosmologically, this contribution is parameterized by the deviation in the effective number of relativistic degrees of freedom, $\Delta N_{\rm eff} \equiv N_{\rm eff} - N_{\rm eff}^{\rm SM}$. 
	If a specific generation $\nu_{Ri}$ attains thermal equilibrium and subsequently decouples while relativistic at a temperature $T_{D,i}$, its relic contribution to $\Delta N_{\rm eff}$ is given by\begin{equation}\Delta N_{\rm eff}^{(i)} \simeq \left[ \frac{g_{*s}(T_{\nu,D})}{g_{*s}(T_{D,i})} \right]^{4/3},\end{equation}where $T_{\nu,D} \approx 2\text{ MeV}$ denotes the active-neutrino decoupling temperature. In the limit where all three right-handed neutrinos thermalize and decouple early in the epoch where $g_{*s} = 106.75$, one obtains the benchmark prediction
	\begin{equation}
		\Delta N_{\rm eff}^{\rm thermal} \simeq 3 \left( \frac{10.75}{106.75} \right)^{4/3} \simeq 0.14.
	\end{equation}
	Conversely, if the effective couplings associated with \(\nu_{Ri}\) are sufficiently feeble, these states never attain thermal equilibrium with the Standard Model plasma. Instead, they can be produced non-thermally through freeze-in processes, contributing to the total radiation energy density and potentially leading to a non-negligible shift in the effective number of relativistic species, \(N_{\rm eff}\).
	~\cite{Luo:2020fdt}.
	These predicted shifts in $N_{\rm eff}$ can be directly tested by cosmological observations. The Standard Model benchmark $N_{\rm eff}^{\rm SM} = 3.045$~\cite{Mangano:2005cc, Grohs:2015tfy, deSalas:2016ztq} aligns well with current limits from Planck 2018 ($N_{\rm eff} = 2.99 \pm 0.17$ at $68\%$ CL)~\cite{Planck:2018vyg} and Big Bang Nucleosynthesis ($2.3 < N_{\rm eff} < 3.4$ at $95\%$ CL)~\cite{Cyburt:2015mya}. Next-generation experiments such as CMB Stage-IV (CMB-S4) will reach an unprecedented sensitivity of $\Delta N_{\rm eff} \approx \pm 0.06$~\cite{Abazajian:2019eic}, making $N_{\rm eff}$ a sensitive probe of Dirac neutrino models and their associated right-handed sectors~\cite{Nanda:2019nqy, Abazajian:2019oqj, FileviezPerez:2019cyn, Han:2020oet, Luo:2020sho, Adshead:2020ekg, Luo:2020fdt, Borah:2020boy, Mahanta:2021plx, Biswas:2021kio, Li:2022yna, Borah:2022obi, Borah:22enh, Biswas:2022fga, Biswas:2022vkq}.
	
	The work is organized as follows: providing the model details, neutrino mass structure in section~\ref{sec:model}, we discussed the freeze-in of DM study in detail in section \ref{sec:FIMP} and their respective sub-sections. The contribution to effective degrees of freedom due to the additional light Dirac neutrinos is discussed in section \ref{neffsec}. Finally, we concluded our work in section \ref{sec:conclusion}. 
	
	\section{The Model}
	\label{sec:model}
	In this section, we present an overview of our model construction in detail.
	The discrete non-abelian symmetry group $A_4$ describes the relevant interactions among the particles and restricts unwanted interactions\footnote{Since the discrete $A_4$ symmetry is spontaneously broken, the possible formation of domain walls requires consideration. We assume either that the $A_4$-breaking occurs before/during inflation with sufficient dilution, or that a small explicit breaking lifts the vacuum degeneracy.}. To achieve this, the particles in the model have specific assignments of the irreducible representation of the group as shown in table \ref{tab1}. An additional $U(1)_L$ charge to avoid unwanted terms as the discrete $A_4$ symmetry group was not sufficient enough. Finally, for the DM stability, we introduced $Z_2$ symmetry, where the dark sector particles remain odd under $Z_2$. 
	
	In this model, we extended the Standard Model with a pair of heavy vector-like neutral fermion $N(=N_L,N_R)$ each being an $A_4$ triplet. To complete the Dirac structure of the proposed model framework, we introduced an $A_4$ triplet light Dirac fermion $\nu_R$. The additional $SU(2)$ singlet scalar fields $\phi$ and $\eta$ are added, which help in neutrino mass generation and the dark sector, respectively. Among the two singlet scalars, $\phi$ being an $A_4$ triplet, will involved in the Dirac mass generation, and $\eta$ being an inert scalar and $Z_2$ odd and will be involved in the DM sector interaction. A Dirac fermion $\chi$ is introduced\footnote{Being a Dirac fermion, we can always write $\chi=\chi_R+ \chi_L$, this will not change anything throughout our analysis therefore we kept it as $\chi$ for simplicity.} with $Z_2$ odd charge to serve as a viable DM candidate. 
	\begin{table}[h]
		\centering
		\begin{tabular}{||c|cccccccc|ccc||}
			\hline
			& $L$ & $e_R$ & $\mu_R$ & $\tau_R$ & $\nu_R$ & $N_L$ & $N_R$ & $\chi$ & $H$ & $\phi$ & $\eta$\\
			\hline
			$SU(2)$&2&1&1&1&1&1&1&1&2&1&1\\
			$A_4$ & 3&1&1$^{\prime \prime}$ &1$^\prime$&3&3&3&1&1&3&3\\
			$Z_2$& +1&+1&+1&+1&+1&+1&+1&-1&+1&+1&-1\\
			$U(1)_L$&1&1&1&1&1&1&1&1&0&0&0\\
			\hline
		\end{tabular}
		\caption{Particles and their charge assignments}
		\label{tab1}
	\end{table}
	
	The interaction Lagrangian for the charged lepton sector is found to be 
	\begin{equation}
		\mathcal{L}_{M_l}=\frac{y_e}{\Lambda}(\overline{L}H\phi)e_R 
		+ \frac{y_\mu}{\Lambda}(\overline{L}\phi)_{1'}H\mu_R
		+\frac{y_\tau}{\Lambda}(\overline{L}\phi)_{1''}H\tau_R+hc.
	\end{equation}
	Here, $\Lambda$ being the cut-off scale mass is introduced to achieve $dim-4$ the Lagrangian for the charged lepton masses. The complete Dirac neutrino Lagrangian is given by:
	\begin{equation}\label{lagNu}
		\mathcal{L}_{D}=
		y_1(\overline{L}N_R)_1 H
		+y_2(\overline{N}_{L} \nu_R)_{3s}\phi
		+y_3(\overline{N}_{L} \nu_R)_{3a}\phi
		+M_N\overline{N}_{L}N_R+hc.
	\end{equation}
	In this Dirac Lagrangian, the first term gives the interaction between the lepton doublet $L$ and the heavy fermion $N_R$ via the SM Higgs $H$. The second and the third terms give the interaction between the light Dirac neutrino and the heavy fermion ($N_L$) via the new scalar $\phi$. The last term generates the heavy neutrino bare mass term through the interaction between heavy neutrino states. 
	
	After electroweak symmetry breaking the VEV alignment of the scalars are given by, $\langle H \rangle = v$ and $\langle \phi \rangle = v_\phi(1,1,1)$. A brief discussion on the scalar potential and their VEV structure is presented in appendix \ref{apen2}. The associated mass matrix for the charged lepton sector is 
	\begin{equation}
		M_l =
		\frac{v}{\Lambda}
		\begin{pmatrix}
			y_e & 0 & 0 \\
			0 & y_\mu & 0 \\
			0 & 0 & y_\tau
		\end{pmatrix}
	\end{equation}
	The Dirac mass matrix, which arises due to the interaction between  $L$ and $N_R$ and this is given by,
	\begin{equation}
		M_D=y_1 v
		\begin{pmatrix}
			1 & 0 &0 \\
			0  &  1&  0\\
			0&0&1
		\end{pmatrix} 
		\label{eq:MD}
	\end{equation}
	Similarly, the corresponding mass matrix due to the interaction of $\nu_R$ and $N_L$ is
	\begin{equation}
		M'_D =v_\phi
		\begin{pmatrix}
			2y_2 & y_3-y_2& -(y_2+y_3)\\
			-(y_2+y_3) & 2y_2 & y_3-y_2\\
			y_3-y_2 & -(y_2+y_3) &2y_2
		\end{pmatrix}
		\label{eq:MDprime}
	\end{equation}
	We also construct a diagonal heavy neutrino mass matrix due to the interaction between $N_L$ and $N_R $, which is given by,
	\begin{equation}
		M_N=M
		\begin{pmatrix}
			1& 0 &0\\
			0&1&0\\
			0&0&1
		\end{pmatrix}    
	\end{equation}
	Thus, the corresponding neutrino mass matrix will be
	\begin{equation}
		m_\nu =-M_D M^{-1}_N M'_D = 
		-\frac{y_1 v v_\phi}{M}
		\begin{pmatrix}
			2y_2 & y_3-y_2& -(y_2+y_3)\\
			-(y_2+y_3) & 2y_2 & y_3-y_2\\
			y_3-y_2 & -(y_2+y_3) &2y_2
		\end{pmatrix}\label{numass}
	\end{equation}
	
	\subsection{Neutrino Yukawa parameter space}
	
	The Dirac neutrino mass matrix \(m_\nu\) depends on the three Yukawa couplings \(y_1,y_2,y_3\), together with the VEVs \(v,v_\phi\) and the right-handed neutrino mass \(M_N\), as given in Eq.~\eqref{numass}. We constrain the model parameter space through a numerical scan over \(y_1,y_2,y_3\). The allowed parameter points are required to be consistent with the latest {\tt NuFit 6.1} neutrino oscillation data. We explored both the  normal hierarchy (NH) and inverted hierarchy (IH) with the lightest neutrino mass zero ($m_1=0$ for NH and $m_3=0$ for IH), and adopt the benchmark values \(v_\phi=100~{\rm GeV}\) and \(M_N=10^{12}~{\rm GeV}\).
	
	The oscillation parameters \(\sin^2\theta_{12}\), \(\sin^2\theta_{13}\), \(\sin^2\theta_{23}\), \(\delta_{\rm CP}\), \(\Delta m^2_{21}\), and \(\Delta m^2_{31}\) are varied within their respective \(3\sigma\) allowed ranges from the global fit~\cite{Esteban:2024eli}. For \(m_1=0\), the remaining neutrino masses are given by $m_2=\sqrt{\Delta m^2_{21}}$ and $m_3=\sqrt{\Delta m^2_{31}}.$
	
	In the basis where the charged-lepton mass matrix is diagonal, the leptonic mixing matrix is identified with \(U_{\rm PMNS}\), which is written in the standard parametrization as
	\begin{equation}
		U_{\rm PMNS} =
		\begin{pmatrix}
			c_{12}c_{13} & s_{12}c_{13} & s_{13}e^{-i\delta_{\rm CP}}\\
			-s_{12}c_{23}-c_{12}s_{23}s_{13}e^{i\delta_{\rm CP}} &
			c_{12}c_{23}-s_{12}s_{23}s_{13}e^{i\delta_{\rm CP}} &
			s_{23}c_{13}\\
			s_{12}s_{23}-c_{12}c_{23}s_{13}e^{i\delta_{\rm CP}} &
			-c_{12}s_{23}-s_{12}c_{23}s_{13}e^{i\delta_{\rm CP}} &
			c_{23}c_{13}
		\end{pmatrix},
	\end{equation}
	where \(c_{ij}=\cos\theta_{ij}\) and \(s_{ij}=\sin\theta_{ij}\).
	
	Since \(m_\nu\) in Eq.~\eqref{numass} is a non-symmetric Dirac mass matrix, it is not directly diagonalized by \(U_{\rm PMNS}\). Instead, the physical neutrino masses and the left-handed mixing matrix are obtained from the Hermitian combination \(m_\nu m_\nu^\dagger\). The matrix reconstructed from the oscillation data is therefore
	\begin{equation}
		{\cal M}_\nu^{2}=U_{\rm PMNS}
		{\rm diag}(m_1^2,m_2^2,m_3^2)
		U_{\rm PMNS}^\dagger .
		\label{eq:Mnu2_reconstructed}
	\end{equation}
	For each scanned parameter point, we calculate the corresponding model prediction \(m_\nu m_\nu^\dagger\) and require it to reproduce the reconstructed matrix in Eq.~\eqref{eq:Mnu2_reconstructed}. Equivalently, the eigenvalues of \(m_\nu m_\nu^\dagger\) must reproduce \(m_1^2,m_2^2,m_3^2\), while its left-handed diagonalization matrix must be consistent with the experimentally allowed \(U_{\rm PMNS}\).
	
	For simplicity, we restrict the Yukawa couplings \(y_1,y_2,y_3\) to be real, corresponding to a CP-conserving benchmark in the model parameters.

	\begin{figure}[h!]
		\centering
		\includegraphics[width=0.45\linewidth]{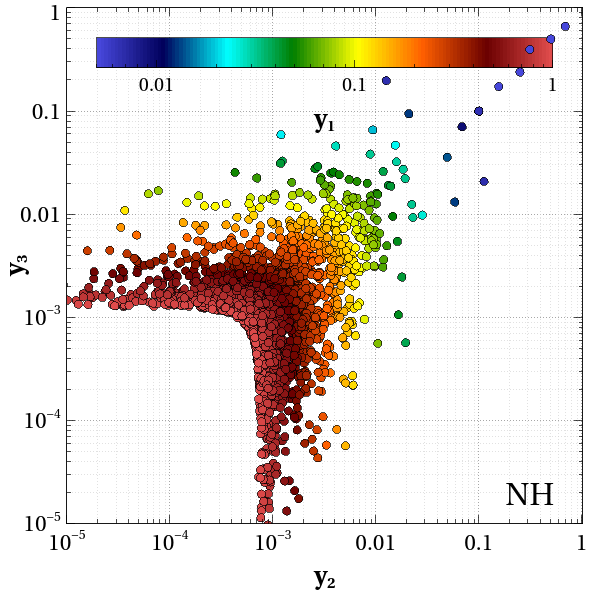}
		\includegraphics[width=0.45\linewidth]{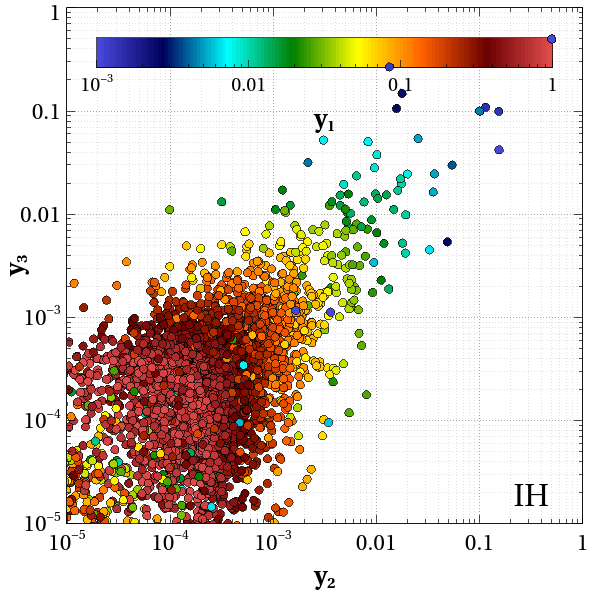}
		\caption{$3\sigma$ allowed parameter space for the Yukawa couplings from the {\tt NuFit 6.1} oscillation parameters. The left panel shows NH (lightest neutrino mass, $m_1=0$) results while the right panel shows IH (lightest neutrino mass, $m_3=0$). }
		\label{nuFig}
	\end{figure}
	Fig~\ref{nuFig} shows the valid points of Yukawa couplings $y_1,y_2,y_3$ which are consistent with the $3\sigma$ range of neutrino oscillation parameters. The allowed points form a distinct strongly correlated region in $y_2-y_3$ plane where both axes are shown in the logarithmic scale and the value of $y_1$ is represented by a color scale. Both normal as well as inverted orderings allowed points are shown in the two panel of the figure, shows a distinct strongly correlated region. The left figure shows a inverse correlation between $y_1$ and the other two couplings for Normal Hierarchy(NH), with $m_1=0$. The larger values of $y_1$($\sim 0.3-1$) are concentrated in the lower left part of the plane where $y_2\sim 10^{-5}-10^{-3}$ and $y_3\sim 10^{-5}-10^{-3}$. For values of $y_1$ in the intermediate range($\sim 0.01-0.3$) the points are in the middle of the distribution with both $y_2$ and $y_3$ $\sim 10^{-2}-10^{-3}$. Similar trend is followed for the smaller values of the Yukawa $y_1$($\lesssim.001$). Thus the values of $y_2$ and $y_3$ both decreases with the increasing value of $y_1$, but the spread of the points and presence of outliers shows that the correlation is not strictly monotonic.
	On the other hand, the right panel of Fig.~\ref{nuFig} shows a inverse correlation between $y_1$ and the other two couplings for Inverted Hierarchy(IH), with $m_3=0$. The allowed points forms a broad, roughly elliptical cloud with most of the points of $y_2$ and $y_3$ concentrated in the range($\sim 10^{-5}$--$10^{-3}$) for larger values of $y_1(\sim0.1-1)$. The cloud has no distinct lower cutoff for $y_2$ and $y_3$ as we have seen in NH. For values $y_2\gtrsim10^{-2}$ the points narrow into a diagonal band with $y_3\simeq y_2$ up to a factor of a few, and $y_3$ lies slightly above $y_2$.
	
	This correlation arises from the matrix structure of $M'_D$ in Eq.~\eqref{eq:MDprime}, where $y_2$ and $y_3$ enter only through the fixed combinations $2y_2$ and $y_3 - y_2$. Since $y_1$ sets the overall Dirac mass scale in Eq.~\eqref{eq:MD}, while $y_2$ and $y_3$ determine the mixing pattern through $M'_D$, satisfying the oscillation data simultaneously requires $y_1$ to trade off against $y_2$ and $y_3$. The finite width of the allowed band reflects the $3\sigma$ spread of the oscillation parameters.

	\section{FIMP dark matter}
	\label{sec:FIMP}
	In the dark sector, we have $Z_2$ odd fermion $\chi$ and the scalar $\eta$. The interaction among them with $N_L$ via the coupling $\lambda$ as:
	\begin{equation}
		\mathcal{L}_{Dark}=\lambda \bar{N_L}\chi\eta+hc.
		\label{eq:Dark}
	\end{equation}
	We considered the mass hierarchy of the order as $M_{N_L}>>M_\eta>M_\chi$, such that $\chi$ remains the lightest stable particle and behaves as a potential DM candidate. To establish the Dirac fermion $\chi$ as a Feebly Intereacting Massive Particle (FIMP), the coupling $\lambda$ is chosen to be extremely small, such that $\chi$ never reaches thermal equilibrium. Initially $\chi$ density in the early universe was negligible, later they are produced from the decay of $N_L$ and $\chi$ never attain thermal equilibrium. We keep track of the situation where its relic abundance is built up gradually through the out-of-equilibrium decay $N_L\rightarrow \chi+\eta$.
	
	The decay width of the process is given by \cite{Hall:2009bx}
	\begin{equation}
		\Gamma_{N_L} \;=\; \frac{\lambda^2 M_{N_L}}{16\pi}
		\left(1 - \frac{m_\chi^2}{M_{N_L}^2}\right)^{2}\,.
		\label{eq:width}
	\end{equation}
	where, we have neglected the $m_\eta$ dependence. The entire  dark sector phenomenology is related to this decay width. We have used this decay width in calculating the relic density, the thermalization condition, and the yield evolution of the DM particle $\chi$.

	The evolution of the $\chi$ number density "$n_\chi$" is governed by the~\cite{Hall:2009bx}
	Boltzmann equation
	\begin{equation}
		\dot n_\chi + 3H n_\chi \;=\;
		\int d\Pi_\chi\, d\Pi_{N_L}\, d\Pi_\eta\,
		(2\pi)^4\, \delta^{(4)}(p_{N_L} - p_\chi - p_\eta)\,
		|\mathcal{M}|^2_{N_L\to\chi\eta}\, f_{N_L}\,,
		\label{eq:boltzmann}
	\end{equation}
	where $H$ is the Hubble rate, $d\Pi_i = d^3p_i/[(2\pi)^3\, 2E_i]$ are the
	phase-space elements, and $f_{N_L}$ is the phase space density of $N_L$. We assume that initial $\chi$ abundance is negligible after
	inflation, the inverse (back-reaction) process is dropped and the production
	is entirely IR-dominated, peaking at temperatures $T \sim M_{N_L}$. By assuming
	the bath particles follow an equilibrium distribution and neglecting Pauli
	blocking/stimulated emission effects, the Boltzmann equation reduces to,
	\begin{equation}
		\dot n_\chi + 3H n_\chi \;\simeq\;
		\frac{g_{N_L}\, M_{N_L}^2\, \Gamma_{N_L}}{2\pi^2}\,
		T\, K_1\!\left(\frac{M_{N_L}}{T}\right),
		\label{eq:collision}
	\end{equation}
	where, $K_1$ is the first-order modified Bessel function of the second kind and
	$g_{N_L}$ provides the internal degrees of freedom of $N_L$. Here \(g_{N_L}\) counts the spin and flavor degrees of freedom of the decaying \(N_L\), while the contribution from the CP-conjugate decay is included explicitly in the total \(\chi+\bar\chi\) abundance.
	
	After rewriting the $n_\chi$ in terms of comoving yield
	$Y_\chi \equiv n_\chi/s$, with $s = 2\pi^2 g_{*S} T^3/45$ the entropy density.
	By using $\dot T \simeq -HT$ and integrating Eq.~\eqref{eq:collision} over
	temperature, the asymptotic freeze-in yield is~\cite{Hall:2009bx}
	\begin{equation}
		Y_\chi \;\simeq\;
		\frac{135\, g_{N_L}}{8\pi^3\,(1.66)\, g_{*S} \sqrt{g_{*\rho}}}\,
		\frac{M_{\rm Pl}\,\Gamma_{N_L}}{M_{N_L}^2}\,,
		\label{eq:yield}
	\end{equation}
	where $g_{*S}$ and $g_{*\rho}$ are the effective numbers of entropy and
	energy degrees of freedom evaluated at $T \sim M_{N_L}$, and $M_{\rm Pl}$ is
	the Planck mass. The yield is independent of the reheating
	temperature, a characteristic feature of IR freeze-in through a renormalizable operator.
	
	The present-day relic abundance follows directly from the yield,
	\begin{equation}
		\Omega_\chi h^2 \;=\; 2.742 \times 10^{8}\,
		\frac{m_\chi}{\text{GeV}}\; Y_\chi\,,
		\label{eq:omega}
	\end{equation}
	where, the numerical factor converts the comoving yield into the present-day
	density parameter. Since $\chi$ is a Dirac fermion, the full relic density
	will have equal contributions from $\chi$ and $\bar\chi$, for symmetric
	production this amounts to an overall factor of two, which we include in the
	numerical scan. By combining Eqs.~\eqref{eq:width}, \eqref{eq:yield} and
	\eqref{eq:omega} and imposing the observed value
	$\Omega_{\rm DM}h^2 \simeq 0.12$ fixes the freeze-in coupling $\lambda$ as a
	function of $m_\chi$. Because  yield, $Y_\chi \propto \lambda^2$ while relic density,
	$\Omega_\chi h^2 \propto m_\chi\, Y_\chi$, a heavier $\chi$ carries more mass
	per state and therefore requires a smaller coupling $\lambda$ to saturate the same relic
	density. Thus a contour of fixed relic abundance
	$\lambda \propto m_\chi^{-1/2}$ should be observed.

	\subsection{Thermalization check}
	\label{sec:thermalization}
	
	To explicitly verify that DM particle $\chi$ never enters thermal equilibrium with the
	bath, we compare the interaction rate $\Gamma_{N_L}$ associated with the decay
	$N_L \to \chi + \eta$ to the Hubble expansion rate $H(T)$, as a function of a dimensionless variable $z = M_{N_L}/T$.
	The thermally averaged production rate remains
	below the Hubble rate throughout the thermal history in freeze-in scenario~\cite{Hall:2009bx}. The thermally averaged
	decay rate of $N_L$ is given by~\cite{Kang:2014cia}:
	\begin{equation}
		\Gamma_{N_L}(T) \;=\; \Gamma_{N_L}\,\frac{K_1(z)}{K_2(z)},
		\label{eq:thermrate}
	\end{equation}
	where, the ratio of Bessel functions accounts for the relativistic time
	dilation averaged over the $N_L$ distribution, while the Hubble rate reads
	\begin{equation}
		H(T) \;=\; 1.66\,\sqrt{g_{*\rho}}\,\frac{T^2}{M_{\rm Pl}}
		\;=\; 1.66\,\sqrt{g_{*\rho}}\,\frac{M_{N_L}^2}{M_{\rm Pl}\, z^2}\,.
		\label{eq:hubble}
	\end{equation}
	The requirement $\Gamma_{N_L}(T)/H(T) < 1$ for all $z$ guarantees that $\chi$
	is a genuine FIMP. Since $\Gamma_{N_L} \propto \lambda^2$, this ratio is
	controlled by the same coupling that fixes the relic density. The values
	of $\lambda$ selected by Eq.~\eqref{eq:omega}, remain many orders of
	magnitude below unity across the entire thermal history.
	\begin{figure}[h!]
		\centering
		\includegraphics[width=0.65\textwidth]{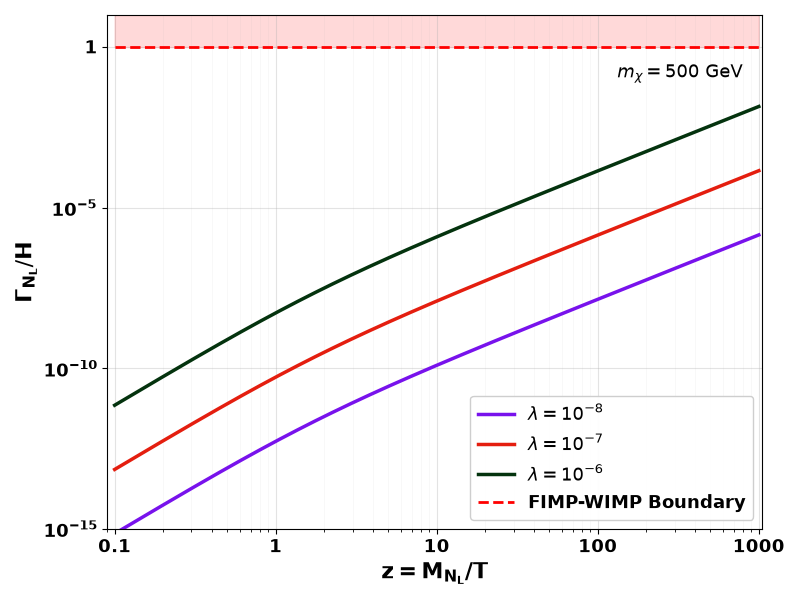}
		\caption{Ratio $\Gamma_{N_L}/H$ as a function of $z = M_{N_L}/T$ for representative values of the freeze-in coupling $\lambda = 10^{-6}, 10^{-7}, 10^{-8}$. The shaded band marks
			$\Gamma_{N_L}/H \gtrsim 1$, the regime in which $\chi$ would thermalize with the bath (WIMP regime).}
		\label{fig:thermalization}
	\end{figure}
	
	For all three benchmark values of $\lambda$ shown in
	Fig.~\ref{fig:thermalization}, $\Gamma_{N_L}/H$ remains below unity across the
	entire range of $z$ probed. Even at late times (($z \sim 10^3$)) interaction rate approaches  but has never entered the shaded WIMP region.
	This confirms that the DM particle $\chi$ is a true FIMP candidate, as it never attains equilibrium with the thermal bath. Its final density is entirely determined by the out-of-equilibrium decay of $N_L$ rather than by a freeze-out condition. As expected with increasing value of $\lambda$ increases the interaction rate, as $\Gamma_{N_L}\propto\lambda^2$. Thus the ratio $\Gamma_{N_L}/H$ increases with increasing $\lambda$, which brings the system closer to
	the thermalization condition $\Gamma_{N_L}/H\gtrsim 1$. This behavior establishes the important role of $\lambda$ in controlling the overall strength of the $N_L$--$\chi$--$\eta$ interaction.
	
	\subsection{Freeze-in Relic density}
	The DM candidate $\chi$ is produced through the decay $N_L \to \chi + \eta$ is 
	mediated by the coupling $\lambda$ in ~\eqref{eq:Dark}. After solving the Boltzmann equation
	for the yield $Y_\chi = n_\chi/s$ and imposing constraint on observed relic density  $\Omega_\chi h^2 \approx 0.12$
	fixes $\lambda$ as a function of the DM mass $m_\chi$.
	\begin{figure}[h!]
		\centering
		\includegraphics[width=0.5\textwidth]{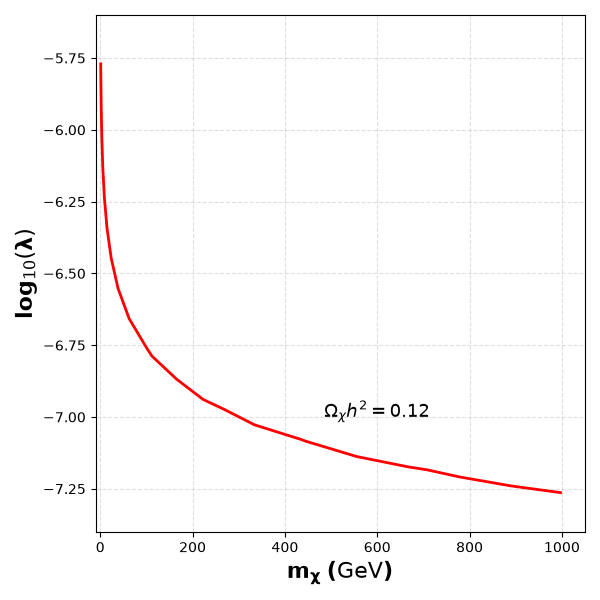}
		\caption{Value of the freeze-in coupling $\log_{10}\lambda$ required to
			reproduce the observed relic abundance $\Omega_{DM}h^2 \approx 0.12$, as a
			function of the dark matter mass $m_\chi$, for $M_{N_L}=10^{12}$ GeV.}
		\label{fig:contour}
	\end{figure}
	Fig.~\ref{fig:contour} shows the coupling required to saturate the observed relic abundance decreases monotonically with $m_\chi$, from
	$\lambda \sim 10^{-5.5}$ at $m_\chi \to 0$ to $\lambda \sim 10^{-7.5}$ at
	$m_\chi = 1000$~GeV. The inverse behavior is a characteristic feature of freeze-in production, where heavier DM particle $\chi$ carries more mass per state, thus, fewer particles are required for yielding the same mass density. Since the final yield varies as $Y_\chi \propto\lambda ^2$, larger $m\chi$ necessitates a smaller value of $\lambda$.
	For the entire mass range $m_\chi[1-1000]$~GeV, the coupling $\lambda$ stays many orders of magnitude below the values required for the thermalization threshold, indicating that the relic abundance in this model is accommodated by the freeze-in rather than freeze-out.
	
\begin{figure}[h!]
		\centering
		\includegraphics[width=0.55\textwidth]{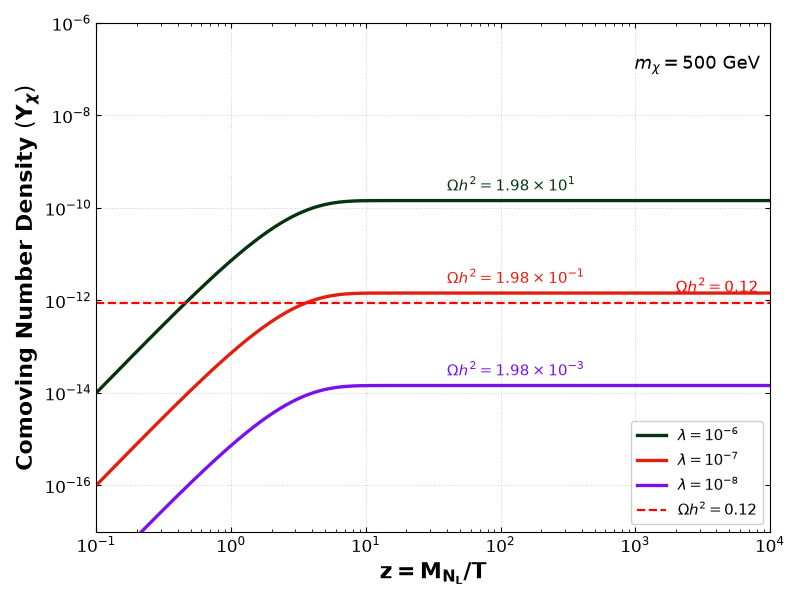}
		\caption{Comoving yield $Y_\chi$ as a function of $z = M_{N_L}/T$ for $\lambda = 10^{-6}, 10^{-7}, 10^{-8}$. The dashed line indicates the yield required to reproduce the observed relic density $\Omega_{DM}h^2 \approx 0.12$.}
		\label{fig:yield}
	\end{figure}

	Finally, we solve the Boltzmann equation for $Y_\chi(z)$ directly to track the
	build-up of the DM abundance for the same three benchmark couplings.
	Fig.~\ref{fig:yield} shows the characteristic evolution of the DM yield in the freeze-in scenario. At early times, corresponds to $z\ll 1$ the abundance of $\chi$ is remain negligible due to the feeble interaction strength. As the universe cools down, the decay $N_L \rightarrow \chi +\eta$ becomes the dominant source of $m_\chi$ which corresponds to a rapid increase in comoving yield ($Y_\chi$). At later times with $z\gg1$, the abundance of ${N_L}$ is Boltzmann is suppressed, and its decay contribution to $m_\chi$ production decreases rapidly. Thus the yield reaches a constant asymptotic value. This saturation behavior at large $z$ is a characteristic of freeze-in. In freeze-out yield curves of a WIMP falls exponentially before plateauing at a much later stage.
	The relative normalization of the three curves follows $Y_\chi \propto \lambda^2$,  with larger value of $\lambda$ resulting into larger final yield. 
	
	\section{Thermalization of light right-handed neutrinos and $\Delta N_{\rm eff}$}\label{neffsec}
	
	Recalling the interaction Lagrangian in Eq.~\eqref{lagNu}, the interactions responsible for populating the light right-handed neutrinos are given by
	\begin{equation}
		\mathcal{L}_{Y}
		\supset
		y_2\left(\overline{N}_{L}\nu_R\right)_{3s}\phi
		+
		y_3\left(\overline{N}_{L}\nu_R\right)_{3a}\phi
		+\mathrm{h.c.},
	\end{equation}
	where the subscripts $3s$ and $3a$ denote the symmetric and antisymmetric $A_4$ contractions, respectively. For the numerical analysis, we consider
	$10^{-4}\lesssim y_2,y_3\lesssim 1$,
	with the heavy fermions having masses of order
	$M_{N_L}\sim10^{12}\ {\rm GeV}$. We assume a reheating temperature much larger than the heavy-fermion mass scale, such that the heavy fermions, $\phi$ and the other relevant particles are initially in thermal equilibrium with the primordial plasma.
	
	It is important to note that the physical couplings relevant for thermalization are the eigenvalues of the full Yukawa matrix rather than $y_2$ and $y_3$ separately. In the present model, the Yukawa matrix can be written schematically as
	\begin{equation}
		Y_{\nu_R}=y_2Y_{3s}+y_3Y_{3a},
	\end{equation}
	where $Y_{3s}$ and $Y_{3a}$ denote the matrices associated with the symmetric and antisymmetric $A_4$ contractions. The effective couplings of the three right-handed neutrino eigenstates are therefore obtained as $Y_i^2=
	{\rm Eigenvalue}\left(Y_{\nu_R}^{\dagger}Y_{\nu_R}\right).$
	
	The dominant production mechanism considered here is the decay and inverse-decay process,
	$N_L\leftrightarrow \nu_{Ri}\phi$.
	The corresponding $2\rightarrow2$ scattering processes, such as
	$\nu_RN_L\leftrightarrow\nu_RN_L$, are subdominant in the parameter region of interest since their rates involve higher powers of the Yukawa couplings, schematically $\Gamma_{\rm scatt}\propto Y_i^4$, whereas the decay and inverse-decay rate scales as $\Gamma_{N_L}\propto Y_i^2$. We therefore neglect the scattering contribution in the following estimate.
	
	For $M_{N_L}\gg m_\phi,m_{\nu_R}$, the decay width is approximately
	\begin{equation}
		\Gamma_{N_L}
		\simeq
		\frac{|Y_i|^2}{16\pi}M_{N_L},
	\end{equation}
	up to possible multiplicity and phase-space factors. The competition between the decay rate and the Hubble expansion can be characterized by
	\begin{equation}
		R_i(T)\equiv
		\frac{\Gamma_{N_L}}{H(T)},
	\end{equation}
	At $T\simeq M_{N_L}$, this ratio gives
	\begin{equation}
		R_i(T\simeq M_{N_L})
		\simeq
		\frac{|Y_i|^2M_{\rm Pl}}
		{16\pi,1.66\sqrt{g_*},M_{N_L}}
		\sim
		10^4|Y_i|^2
	\end{equation}
	for $M_{N_L}\sim10^{12}$ GeV and $g_*\simeq106.75$. Thus, Yukawa couplings of order $10^{-2}$ or larger can make the decay rate comparable to or larger than the Hubble expansion rate around $T\sim M_{N_L}$, indicating the possibility of efficient thermalization. For smaller couplings, the production becomes increasingly inefficient and the right-handed neutrinos can remain out of equilibrium.
	
	Rather than introducing an effective thermalization fraction, we determine the right-handed neutrino abundance by numerically solving the Boltzmann equation for each $\nu_{Ri}$ state. Defining $z\equiv\frac{M_{N_L}}{T},$
	the abundance $Y_{\nu_{Ri}}\equiv n_{\nu_{Ri}}/s$ evolves according to~\cite{Luo:2020fdt}
	\begin{equation}
		\frac{dY_{\nu_{Ri}}}{dz}=\frac{\Gamma_{N_L}}
		{H(z)z}
		\frac{K_1(z)}{K_2(z)}
		Y_{N_L}^{\rm eq}(z)
		\left(
		1-\frac{Y_{\nu_{Ri}}}{Y_{\nu_R}^{\rm eq}}
		\right),
		\label{eq:nuR_Boltzmann}
	\end{equation}
	where $K_{1,2}$ are modified Bessel functions of the second kind and $Y_{N_L}^{\rm eq}(z)=\frac{45g_{N_L}}{4\pi^4g_{*s}}
	z^2K_2(z)$ as the equilibrium abundance of the heavy fermion. The equilibrium abundance of one relativistic right-handed neutrino degree of freedom is
	\begin{equation}
		Y_{\nu_R}^{\rm eq}=\frac{135\zeta(3)}{8\pi^4}
		\frac{g_{\nu_R}}{g_{*s}}.
	\end{equation}
	The factor $K_1(z)/K_2(z)$ accounts for the thermal time-dilation correction to the decay rate.
	
	Eq.~\eqref{eq:nuR_Boltzmann} continuously describes both the freeze-in and thermalization regimes. For sufficiently small $Y_i$, the abundance remains much smaller than its equilibrium value and the production term is proportional to $Y_i^2$, leading to the expected freeze-in behavior. For sufficiently large $Y_i$, the inverse-decay term becomes important and drives the abundance towards $Y_{\nu_R}^{\rm eq}$, corresponding to thermalization. Therefore, no phenomenological interpolation between the two regimes is required.
	
	Once the heavy states have disappeared from the thermal bath, the right-handed neutrinos decouple from the primordial plasma. Their contribution to the radiation density can then be expressed in terms of their final abundance as
	\begin{equation}
		\Delta N_{\rm eff}^{(i)}
		\simeq
		\frac{Y_{\nu_{Ri}}^{\rm final}}
		{Y_{\nu_R}^{\rm eq}}
		\left[
		\frac{g_{*s}(T_{\nu,D})}
		{g_{*s}(T_{D,i})}
		\right]^{4/3},
		\label{eq:DeltaNeff_nuR}
	\end{equation}
	where $T_{\nu,D}$ denotes the active-neutrino decoupling temperature, for which $g_{*s}(T_{\nu,D})\simeq10.75$. The total contribution from the three right-handed neutrinos is therefore
	\begin{equation}
		\Delta N_{\rm eff}^{\nu_R}=
		\sum_{i=1}^{3}
		\Delta N_{\rm eff}^{(i)}.
		\label{eq:DeltaNeff_total}
	\end{equation}
	In the fully thermalized limit, $Y_{\nu_{Ri}}^{\rm final}\rightarrow Y_{\nu_R}^{\rm eq}$, and for decoupling at $g_{*s}(T_{D,i})=106.75$ one obtains
	\begin{equation}
		\Delta N_{\rm eff}^{(i)}
		\simeq
		\left(\frac{10.75}{106.75}\right)^{4/3}
		\simeq0.046,
	\end{equation}
	and consequently
	\begin{equation}
		\Delta N_{\rm eff}^{\nu_R}
		\simeq0.14
	\end{equation}
	for three thermalized right-handed neutrino species.
	
	To distinguish the thermalized and non-thermal regimes, we define the final abundance ratio
	\begin{equation}\label{rieq}
		r_i\equiv
		\frac{Y_{\nu_{Ri}}^{\rm final}}
		{Y_{\nu_R}^{\rm eq}},
	\end{equation}
	where $Y_{\nu_{Ri}}^{\rm final}$ is obtained by numerically solving Eq.~\eqref{eq:nuR_Boltzmann}, while $Y_{\nu_R}^{\rm eq}$ denotes the equilibrium abundance of a relativistic right-handed neutrino. For $r_i\simeq1$, the $\nu_{Ri}$ abundance approaches its equilibrium value and the state is considered thermalized, whereas $r_i\ll1$ indicates inefficient production and a non-thermal (freeze-in) population.
	
	\begin{figure}[h!]
		\centering
		\includegraphics[width=0.65\linewidth]{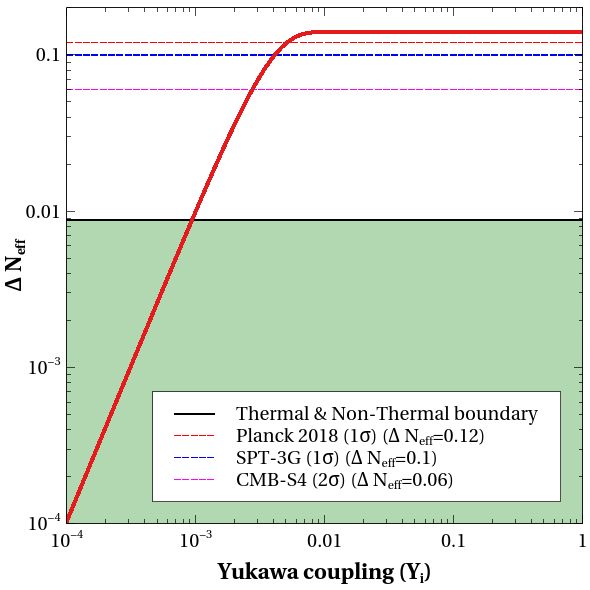}
		\caption{The variation of $\Delta N_{\rm eff}$ vs. the Yukawa coupling $Y_i$. The black line separating the white and green shaded region as thermal and non-thermal contributions to $\Delta N_{\rm eff}$. The respective experimental sensitivities are shown by the horizontal dashed lines and are mentioned in the key text.   }
		\label{figNeff}
	\end{figure}
	For the parameter points obtained from the neutrino-mass scan, we evaluate the physical Yukawa eigenvalues $Y_i$ and solve Eq.~\eqref{eq:nuR_Boltzmann} numerically for each state. This procedure directly determines the corresponding $\nu_R$ abundances and their contribution to $\Delta N_{\rm eff}$, thereby accounting for the transition between inefficient production and thermalization without introducing an additional phenomenological thermalization parameter.
	
	Fig. \ref{figNeff}, shows the variation of $\Delta N_{\rm eff}$ as a function of the Yukawa coupling distinguishing both thermal and non-thermal regions. In our numerical analysis, we use $r_i\geq0.99$ to identify effectively thermalized states, and the rest of the regions are non-thermal. The black solid horizontal line separates the thermal and non-thermal regions evaluated using \eqref{rieq}. These numerical thresholds are used only to classify the scan points and do not represent sharp physical boundaries between the two regimes \cite{Luo:2020fdt,Mahapatra:2023oyh}. In the benchmark parameter region considered here, we find a maximum characteristic prediction of: $\Delta N_{\rm eff} \simeq 0.14$.
	This shift falls comfortably within the projected sensitivities of next-generation cosmic microwave background experiments, including CMB-S4~\cite{Abazajian:2019eic} and CMB-HD~\cite{CMB-HD:2022bsz}. 
	
	\section{Conclusion}
	\label{sec:conclusion}
	
	In this work we have proposed a unified framework in which the smallness of the Dirac neutrino mass and the observed dark matter relic abundance both originate from feeble interactions in an extended particle content governed by an $A_4 \times Z_2 \times U(1)_L$ symmetry. A heavy pair of vector-like fermions $N_R$ and $N_L$, both triplet under $A_4$ flavor symmetry, and the flavon fields $\phi, \eta$ were introduced. The $A_4$ symmetry is spontaneously broken by the flavon VEVs to generate the Dirac neutrino mass matrix through a seesaw-like structure as given by Eq.~(9). By further imposing a $Z_2$ symmetry under which $\chi$ and $\eta$ are odd, ensures the lighter of the two states, $\chi$, absolutely stable, providing a natural dark matter candidate produced non thermally through the decay $N_L \to \chi + \eta$.
	
	A numerical scan of the Dirac sector Yukawa couplings against the $3\sigma$ constraint from global fit neutrino oscillation data shows a well-defined, correlated region of the Dirac sector in both the mass ordering patterns. In the FIMP DM section, after solving the Boltzmann equation for the dark sector, it shows that a coupling in the range of $\lambda \sim 10^{-5.5}$--$10^{-7.5}$ reproduces the
	observed relic density $\Omega_{DM}h^2 \approx 0.12$ over
	$m_\chi \in [1,1000]$~GeV. Also these Yukawa couplings in this range keep the ratio $\Gamma_{N_L}/H$ below unity throughout the thermal history. This has confirmed that $\chi$ is a potential candidate for Feebly Interacting Massive Particle (FIMP) rather than a thermal WIMP. The corresponding yield evolution plot shows the characteristic freeze-in profile with negligible abundance at early times, a rapid increase once $N_L$ becomes non-relativistic, and saturation to a constant late time value.
	
	The light Dirac fermions added in the model to generate neutrino mass also play a crucial role in addressing deviation from the effective number of neutrino species. The interaction term $N_L\nu_R\phi$ allows scattering, decay, and inverse decay processes, and among them the decay process dominates over the two. With the choice of Yukawa couplings, the decay process dictates whether the Dirac neutrinos $\nu_R$ contributes thermally or non-thermally to the $N_{\rm eff}$. With the chosen parameter space, we quantitatively estimate the total contribution coming from the thermal and non-thermal scenarios and distinguish their contributions. The maximum obtained value $\Delta N_{\rm eff}\sim0.14$ falls comfortably within the next-gen experimental searches. The resulting cosmological contribution provides a direct connection between the neutrino-mass sector and early-Universe cosmology. In particular, the scalar $\phi$, whose vacuum expectation value determines the effective Dirac neutrino mass scale, also controls the interactions responsible for populating the right-handed neutrinos. Consequently, the parameter space associated with $v_\phi$ and the Yukawa couplings is constrained simultaneously by the light-neutrino masses and the cosmological requirement on $\Delta N_{\rm eff}$.
	
	In a final note, this work establishes a common heavy-mediator and symmetry framework for Dirac neutrino masses and FIMP dark matter. The heavy vector-like fermion simultaneously mediates the Dirac neutrino mass generation and the freeze-in production of dark matter, while the flavon VEV controls the light-neutrino mass scale and flavour structure. 
	
	\section*{Acknowledgement}
	The authors acknowledge the use of the open-source AI tool as a writing assistant for manuscript language editing and as a secondary check for algebraic consistency.
	\appendix
	\section{$A_4$ group and product rules}\label{a4p}
	$A_4$, the symmetry group of a regular tetrahedron, is a discrete non-Abelian group consisting of the $12$ even permutations of four objects. It possesses four irreducible representations: three one-dimensional representations denoted by $\mathbf{1}$, $\mathbf{1'}$, and $\mathbf{1''}$, and one three-dimensional representation denoted by $\mathbf{3}$. 
	
	The group $A_4$ can be generated by two basic generators, $S$ and $T$, which satisfy the presentation relations:
	\begin{equation}
		S^2 = T^3 = (ST)^3 = I,
	\end{equation}
	where $I$ is the identity element. In a diagonal $T$ basiswhich conveniently renders the charged lepton mass matrix diagonal. The group generators in the triplet representation $\mathbf{3}$ can be represented as:
	\begin{equation}
		S = \frac{1}{3} \begin{pmatrix} -1 & 2 & 2 \\ 2 & -1 & 2 \\ 2 & 2 & -1 \end{pmatrix}, \quad 
		T = \begin{pmatrix} 1 & 0 & 0 \\ 0 & \omega & 0 \\ 0 & 0 & \omega^2 \end{pmatrix},
	\end{equation}
	where $\omega = e^{i 2\pi / 3} = -1/2 + i\sqrt{3}/2$ is the complex cube root of unity.
	
	The tensor product rules for the singlet representations are given by~\cite{Ishimori:2010au}:
	\begin{equation*}
		\begin{split}
			\mathbf{1} \otimes \mathbf{1} = \mathbf{1}, \quad 
			\mathbf{1'} \otimes \mathbf{1'} = \mathbf{1''}, \quad 
			\mathbf{1'} \otimes \mathbf{1''} = \mathbf{1}, \quad 
			\mathbf{1''} \otimes \mathbf{1''} = \mathbf{1'}.
		\end{split}
	\end{equation*}
	
	The direct product of two triplet representations decomposes into the direct sum:
	\begin{equation*}
		\mathbf{3} \otimes \mathbf{3} = \mathbf{1} \oplus \mathbf{1'} \oplus \mathbf{1''} \oplus \mathbf{3}_s \oplus \mathbf{3}_a,
	\end{equation*}
	where the subscripts $s$ and $a$ denote the symmetric and anti-symmetric triplet combinations, respectively. For two triplets $a = (a_1, a_2, a_3)^T$ and $b = (b_1, b_2, b_3)^T$, the explicit decomposition into irreps in the $T$-diagonal basis is expressed as~\cite{Altarelli:2005yp}:
	\begin{equation}\label{a4r}
		\begin{split}
			\mathbf{1}   &\sim a_1 b_1 + a_2 b_3 + a_3 b_2, \\
			\mathbf{1'}  &\sim a_3 b_3 + a_2 b_1 + a_1 b_2, \\
			\mathbf{1''} &\sim a_2 b_2 + a_1 b_3 + a_3 b_1, \\
			\mathbf{3}_s &\sim (2 a_1 b_1 - a_2 b_3 - a_3 b_2,\, 2 a_3 b_3 - a_1 b_2 - a_2 b_1,\, 2 a_2 b_2 - a_1 b_3 - a_3 b_1)^T, \\
			\mathbf{3}_a &\sim (a_2 b_3 - a_3 b_2,\, a_1 b_2 - a_2 b_1,\, a_3 b_1 - a_1 b_3)^T.
		\end{split}
	\end{equation}
	\section{Scalar potential and vacuum structure}\label{apen2}
	The scalar sector of the model consists of the Standard Model Higgs doublet \(H\), an \(A_4\)-triplet flavon \(\phi\), and an \(A_4\)-triplet inert scalar \(\eta\). The fields \(\phi\) and \(\eta\) are singlets under the SM gauge group, while \(\eta\) is odd under the imposed \(Z_2\) symmetry. The \(Z_2\) parity forbids terms containing an odd number of \(\eta\) fields, and we assume the parameter region in which the \(Z_2\)-preserving vacuum \(\langle\eta\rangle=0\) is realized.
	
	A general renormalizable scalar potential relevant for the vacuum structure can be written as
	\begin{equation}
		V=V_H+V_\phi+V_\eta+V_{H\phi}+V_{H\eta}+V_{\phi\eta},    
	\end{equation}
	where
	\begin{equation}
		V_H=
		-\mu_H^2 H^\dagger H
		+\lambda_H(H^\dagger H)^2 ,    
	\end{equation}
	and the \(A_4\)-invariant potentials involving the two triplet scalars can be expressed in terms of their irreducible \(A_4\) contractions. In particular,
	\begin{equation}
		\begin{aligned}
			V_\phi={}&
			\mu_\phi^2(\phi^\dagger\phi)_{\mathbf1}
			+\lambda_{\phi1}
			\left[(\phi^\dagger\phi)_{\mathbf1}\right]^2
			+\lambda_{\phi2}
			(\phi^\dagger\phi)_{\mathbf1'}
			(\phi^\dagger\phi)_{\mathbf1''}
			\\
			&+
			\lambda_{\phi3}
			\left[(\phi^\dagger\phi)_{\mathbf3_s}
			\cdot
			(\phi^\dagger\phi)_{\mathbf3_s}\right]_{\mathbf1}
			+
			\lambda_{\phi4}
			\left[(\phi^\dagger\phi)_{\mathbf3_a}
			\cdot
			(\phi^\dagger\phi)_{\mathbf3_a}\right]_{\mathbf1}.
		\end{aligned}
	\end{equation}
	Similarly, the \(Z_2\)-odd scalar \(\eta\) has
	\begin{equation}
		\begin{aligned}
			V_\eta={}&
			\mu_\eta^2(\eta^\dagger\eta)_{\mathbf1}
			+\lambda_{\eta1}
			\left[(\eta^\dagger\eta)_{\mathbf1}\right]^2
			+\lambda_{\eta2}
			(\eta^\dagger\eta)_{\mathbf1'}
			(\eta^\dagger\eta)_{\mathbf1''}
			\\
			&+
			\lambda_{\eta3}
			\left[(\eta^\dagger\eta)_{\mathbf3_s}
			\cdot
			(\eta^\dagger\eta)_{\mathbf3_s}\right]_{\mathbf1}
			+
			\lambda_{\eta4}
			\left[(\eta^\dagger\eta)_{\mathbf3_a}
			\cdot
			(\eta^\dagger\eta)_{\mathbf3_a}\right]_{\mathbf1}.
		\end{aligned}    
	\end{equation}
	The interactions between the SM Higgs and the \(A_4\)-triplet scalars are
	\begin{equation}
		V_{H\phi}
		=
		\lambda_{H\phi}
		(H^\dagger H)(\phi^\dagger\phi)_{\mathbf1} \quad \text{and } \quad    V_{H\eta}=\lambda_{H\eta}
		(H^\dagger H)(\eta^\dagger\eta)_{\mathbf1}.
	\end{equation}
	The interaction between the two \(A_4\) triplets contains several independent \(A_4\)-invariant contractions and can be parameterized as
	\begin{equation}
		\begin{aligned}
			V_{\phi\eta}={}&
			\lambda_{\phi\eta1}
			(\phi^\dagger\phi)_{\mathbf1}
			(\eta^\dagger\eta)_{\mathbf1}+\lambda_{\phi\eta2}
			(\phi^\dagger\phi)_{\mathbf1'}
			(\eta^\dagger\eta)_{\mathbf1''}+\lambda_{\phi\eta3}
			(\phi^\dagger\phi)_{\mathbf1''}
			(\eta^\dagger\eta)_{\mathbf1'}+\lambda_{\phi\eta4}
			\left[
			(\phi^\dagger\phi)_{\mathbf3_s}
			\cdot
			(\eta^\dagger\eta)_{\mathbf3_s}
			\right]_{\mathbf1}
			\\
			&+
			\lambda_{\phi\eta5}
			\left[
			(\phi^\dagger\phi)_{\mathbf3_a}
			\cdot
			(\eta^\dagger\eta)_{\mathbf3_a}
			\right]_{\mathbf1}
			+
			\lambda_{\phi\eta6}
			\left[
			(\phi^\dagger\eta)_{\mathbf3_s}
			\cdot
			(\eta^\dagger\phi)_{\mathbf3_s}
			\right]_{\mathbf1}
			+
			\lambda_{\phi\eta7}
			\left[
			(\phi^\dagger\eta)_{\mathbf3_a}
			\cdot
			(\eta^\dagger\phi)_{\mathbf3_a}
			\right]_{\mathbf1}.
		\end{aligned}    
	\end{equation}
	Here the subscripts \(\mathbf1,\mathbf1',\mathbf1''\), \(\mathbf3_s\), and \(\mathbf3_a\) denote the corresponding \(A_4\) contractions. Details of the product rule under $A_4$ are presented in the appendix \ref{a4p}.
	
	The scalar sector develops the vacuum structure \(\langle H\rangle=v\) and \(\langle\phi\rangle=v_\phi(1,1,1)\) after potential minimization~\cite{Das:2018qyt}, while the \(Z_2\)-odd scalar \(\eta\) does not acquire a vacuum expectation value, i.e. \(\langle\eta\rangle=0\). Since both \(\eta\) and \(\chi\) are \(Z_2\)-odd, we consider the mass hierarchy \(M_{N_L}\gg m_\eta>m_\chi\), such that \(\eta\) is not stable. In particular, the interactions involving \(N_L\), together with the \(N_L\) mass term and the coupling of \(N_R\) to the SM lepton doublet, induce an effective interaction between \(\eta\), \(\chi\) and the SM neutrino after the heavy neutral fermions are integrated out. Consequently, the heavier \(Z_2\)-odd scalar can decay into the lighter \(Z_2\)-odd fermion and a neutrino, \(\eta\rightarrow\chi+\nu\). Thus, \(\chi\) remains the stable dark matter candidate as the lightest \(Z_2\)-odd particle, while \(\eta\) serves as an intermediate state in the freeze-in production mechanism. A detailed discussion on the scalar sector and the fate of $\eta$ are left for our future works.  
	
	\bibliographystyle{JHEP}
	\bibliography{ref_gpt}
	
\end{document}